\documentclass[aps,pra,onecolumn,groupedaddress,showpacs]{revtex4}
\usepackage{amsfonts}
\usepackage{amssymb}
\usepackage{graphicx}
\usepackage{mathcomp}
\usepackage{amsfonts}
\usepackage{amssymb}
\usepackage{epsfig}

\begin{document}

\title{Deterministic Rendering of BB84 for Practical Quantum
Cryptography\\}

\author{\vspace{0.5cm} M. Lucamarini} \email{marco.lucamarini@unicam.it}
\affiliation{Dipartimento di Fisica, Universit\`a di Camerino, via
Madonna delle Carceri, 9\\ 62032 Camerino, Italy\\}

\author{J. S. Shaari}
\affiliation{Faculty of Science, International Islamic University
of Malaysia (IIUM), P.O Box 141, 25710 Kuantan, Pahang Darul
Makmur, Malaysia\\}

\author{M. R. B. Wahiddin}
\affiliation{Faculty of Science, International Islamic University
of Malaysia (IIUM), P.O Box 141, 25710 Kuantan, Pahang Darul
Makmur, Malaysia\\} \affiliation{Information Security Cluster,
MIMOS Berhad, Technology Park Malaysia, 57000 Kuala Lumpur,
Malaysia}

\begin{abstract}
\vspace{0.5cm} We describe how to modify the BB84 protocol for
quantum cryptography in order to make it deterministic. We study
both theoretical and experimental aspects of this issue, showing
that the new scheme is as secure as the old one, more efficient on
small-scale distances, and within the range of current technology.
\end{abstract}


\maketitle

\section*{Introduction}

Classically it is possible to convey information from one user
(Alice) to another distant user (Bob) with perfect fidelity. This
reliable transmission of information is usually obtained by means
of redundancy, i.e. by repeating the main signal as many times as
necessary for compensating the noise and the losses of the channel
connecting the users. However things turn difficult if Alice and
Bob wish to communicate privately, i.e. keeping any third party
(usually called ``Eve'') ignorant about the information exchanged
over the channel. In this situation redundancy plays against
privacy and is not at all simple to guarantee the confidentiality
of the communication.

Quantum mechanics provide a solution to this issue: by using
quantum signals a secret random key can be distributed at distant
places and eventually used for a private communication. Several
protocols for quantum key distribution (QKD) have been suggested
so far
\cite{BB84,B92,Ekert91,BBM92,Bruss98,BPG99,DPS02,Pho00,SARG04,Bruss03}.
Among these the BB84~\cite{BB84} is arguably the most popular, and
has been implemented both in free-space and in optical fiber
setups even beyond the limit of 100
km~\cite{Ursin2007,Gob04,His06}. Its easiness of implementation
also triggered the commercialization of prototypes by a number of
brands~\cite{QKDcommercial}.

The BB84, as the vast majority of the QKD schemes, is a
\textit{non-deterministic} protocol. A protocol is defined
``deterministic'' when Alice can \textit{in-principle} transmit a
predetermined sequence of bits to Bob in a reliable way. The
``in-principle'' conditions basically mean ``under ideal working
conditions'', e.g. with a noiseless and lossless channel
connecting Alice and Bob, or with a Bob endowed with perfect
detectors. The simplest example of a deterministic protocol is any
protocol working with classical signals (which is reliable but not
secure, as explained above), like a standard fax transmission. On
the contrary it is easy to see why the standard BB84 is
non-deterministic. In the BB84 Alice (Bob) prepares (measures) the
quantum signals using one out of two non-commuting observables, or
bases. Only when the bases chosen by Alice and Bob coincide the
two users get correlated results. In all the other instances they
must discard their bits. This happens with probability $1/2$, and
leads to the waste of half of the acquired bits, on the average.
Then Alice can not, even in principle, transmit a predetermined
sequence of bits to Bob, because the final sequence will depend
also on Bob's choice of the basis, about which Alice has no
control.

Quite recently several deterministic protocols have been
proposed~\cite{Beige02,Goldenberg1995,Reid2000,Long2002,Bostrom02,Yuen2003,Cai2004,Luc05,Deng2003,Deng2004,Shaari06}.
All of them satisfy the requisite of an in-principle reliable
communication, but none of them comes with a rigorous proof of its
unconditional security. The reason is that the newly proposed
deterministic schemes are usually quite different from the
existing ones, thus preventing a straightforward application of
the mathematical tools developed for the standard protocols in the
last two decades. The lack of a security proof in turn prevents a
direct comparison between a deterministic and a non-deterministic
protocol.

The deterministic rendering of BB84 (Det-BB84 hereafter) fills
this gap, and lets a first precise quantification of the
advantages coming from determinism. The point is that the security
of Det-BB84 does not need to be demonstrated \textit{ab initio},
because it follows from the standard BB84's one. This allows us to
directly compare the two protocols, and establish that determinism
increases the rate of secure transmission when Alice and Bob are
separated by a small-scale distance.

The paper is organized as follows. In Section~\ref{sec:BB84} we
review what is known on the possibility of a deterministic
rendering of BB84. In Section~\ref{sec:detBB84} we give our
protocol and discuss its security. Finally in
Section~\ref{sec:practice} we propose a practical implementation
of our protocol and contrast it with the traditional BB84.

\section{Deterministic BB84} \label{sec:BB84}

The possibility of a Det-BB84 is a common
knowledge~\cite{note1_DetBB84}. The very first formulation of BB84
foresees that Bob is endowed with a quantum memory to store the
quantum systems (qubits) received by Alice until Alice's public
disclosure of the bases. After knowing the bases of all the qubits
Bob can measure them in a deterministic way, i.e. without risk of
choosing a wrong basis that makes his measure's outcome random.

However a simple-minded Det-BB84 appears to be totally unfeasible
because of the presence of a quantum memory and of the demanding
security requirements pertaining to QKD. For this reason it has
been used so far only as a mathematical tool apt to simplify the
analysis of the security. Even recently it has been adopted
in~\cite{Bih05} to provide a stronger security for a QKD performed
with BB84.

To better explain the current impracticality of the deterministic
scheme we start our discussion with a basic version of Det-BB84,
reported below, which exploits a quantum memory for the storage of
the qubits. This version of Det-BB84 follows the steps of the BB84
described in~\cite{Bih05}, to which we refer for a detailed
description, and of that reported in~\cite{NC00}. We also refer
to~\cite{Gisin02} and~\cite{Dusek06} for QKD-related technical
details.

\medskip

\noindent \textbf{DET-BB84 (basic version)}

\smallskip

\noindent (1)~\textit{Data-bit choice}. Alice chooses a random
$W$-bit string $d$ (data string), where $W=(4+\eta_c+\eta_m)N$.
The factor $\eta_c$ accounts for the losses of the channel while
$\eta_m$ accounts for the losses of Bob's imperfect quantum
memory.

\noindent (2)~\textit{Basis-bit choice and encoding}. Alice
chooses a random $W$-bit string $b$ (basis string). She encodes
each bit of $d$ on the qubits as $\left\{ \left\vert
0\right\rangle ,\left\vert 1\right\rangle \right\} $ if the
corresponding bit of $b$ is $0$ ($Z$ basis) or $\left\{ \left\vert
+\right\rangle ,\left\vert -\right\rangle \right\} $ if the
corresponding bit of $b$ is $1$ ($X$ basis). Alice sends the
resulting states to Bob.

\noindent (3)~\textit{Storage}. Bob receives on average
$(4+\eta_m)N$ qubits and stores them in a (imperfect) quantum
memory.

\noindent (4)~\textit{Receipt}. He announces the completion of
step \underline{3} on the (authenticated or unjammable) classical
channel.

\noindent (5)~\textit{Basis revelation}. Alice announces $b$.

\noindent (6)~\textit{Deterministic measurement}. Bob retrieves on
average $4N$ qubits from the memory and measures each of them in
the $X$ or $Z$ basis according to the disclosed value of $b$. In
this way the outcome of his measure is deterministic and Alice and
Bob do not discard any bits. After the public announcement by Bob
of the addresses of the lost qubits, Alice and Bob will get on the
average $4N$ pairs of correlated bit, a fraction of which contains
errors due to the possible noise on the channel.

\noindent (7)~Alice selects a subset of $2N$ bits that will serve
as a check on Eve's interference, and tells Bob which bits she
selected.

\noindent (8)~Alice and Bob announce and compare the values of the
$2N$ check bits. If more than an acceptable number disagree, they
abort the transmission.

\noindent (9)~Alice and Bob perform error correction and privacy
amplification on the remaining $2N$ bits to obtain $2M$ private
key bits ($M \leq N$).

\medskip

Points \underline{3}, \underline{5} and \underline{6} of the above
protocol makes it deterministic, because they let Bob always
measure in the right basis. In case of a noiseless and lossless
channel between the users, the newly acquired determinism would
enable the possibility of a \textit{direct
communication}~\cite{Beige02,Bostrom02} via BB84. However, even
for an imperfect channel and quantum memory, one can notice that
the coefficient in front of $M$, the final number of distilled
bits, is 2. This should be compared with the coefficient 1
pertaining to BB84~\cite{NC00}. It is apparent that this
corresponds to a doubling of the theoretical final secure
bit-rate.

A crucial point that makes Det-BB84 as secure as the original
BB84~\cite{Bih05} is the \underline{4}, which represents the
receipt by Bob of the qubits sent by Alice. Without it there's a
risk that Eve delays the qubits until the public disclosure of the
basis, thus gaining for herself the possibility of a deterministic
measurement. In such a case Eve would go entirely undetected.

This security issue apart, point \underline{4} represents the main
obstacle toward a practical implementation of Det-BB84. In fact,
to send a receipt, Bob must acknowledge that a given number of
signals (for example photons) entered his station. The only way to
do that without altering the information carried by the photons is
represented by an ideal quantum nondemolition measurement (QND),
which is still a demanding technology~(\cite{QND}, and references
therein).

%
%
%
Point \underline{4} also implies that Bob must store the qubit
until Alice's basis revelation (point \underline{3}). If we follow
the qubit in its travel we see that the minimum storage time for a
Det-BB84 with a receipt's transmission is $ 2 \tau$, where $\tau$
is the time for a signal to cover the distance between Alice and
Bob: one $\tau$ is to let Bob's receipt reach Alice, and one
$\tau$ is to let Alice transmit the basis to Bob (we assume for
simplicity that Alice and Bob use the same channel, hence the two
times are equal in both directions).

The simplest example of a quantum memory is an optical fiber loop
of length $L$ that allows to store a photon for a time $nL/c$,
with $n$ the refractive index of the fiber, $c$ the velocity of
light in vacuum. At least in this simple case it is plain that the
longer the photons are stored, the lower the probability to
recover. Then, it would be necessary to keep the storage time as
lower as possible.

\section{Practicality and Security of Det-BB84} \label{sec:detBB84}

The considerations of Section~\ref{sec:BB84} suggest that the
impracticality of Det-BB84 is mainly related to its point
\underline{4} (receipt of the qubits). In the following we show
how to remove this point from the protocol without affecting its
security. Despite some steps might result unusual (clock
synchronization, initial measurement of the time delay), they have
already been considered elsewhere and belong to the standard
implicit structure of any QKD.

\medskip

\noindent \textbf{DET-BB84 (practical version)}

\smallskip

\noindent (1)~\textit{Preliminaries}. Alice and Bob measure the
time $\tau$ that a classical pulse (e.g. an intense laser pulse)
employs to cover the distance between them. Then they use the
(authenticated or unjammable) classical channel to (i) publicly
declare the measured time $\tau$ (ii) establish the value of a
positive security parameter, $\Delta$, used later for the security
analysis.

\noindent (2)~\textit{Data-bit choice}. Alice chooses a random
$W$-bit string $d$ (data string), with $W=(4+\eta_c+\eta_m)N$. We
indicate with $d_{i}$ ($i=1,...,W$) the $i$-th bit of the string
$d$.

\noindent (3)~\textit{Basis-bit choice}. Alice chooses a random
$W$-bit string $b$ (basis string). We indicate with $ b_{i}$
($i=1,...,W$) the $i$-th bit of the string $b$.

\noindent (4)~\textit{Encoding and transmission of quantum
information}. Beginning with $i=1$ Alice encodes the data $d_{i}$
into the qubit $q_i$. She encodes each bit of $d$ as $\left\{
\left\vert 0\right\rangle ,\left\vert 1\right\rangle \right\} $ if
the corresponding bit of $b$ is $0$ ($Z$ basis) or $\left\{
\left\vert +\right\rangle ,\left\vert -\right\rangle \right\} $ if
the corresponding bit of $b$ is $1$ ($X$ basis). At time $t^{q}_1$
Alice starts the transmission of the qubits to Bob. At the generic
time $t^{q}_i$ Alice will send out the qubit $q_i$. We note that
the times $t^{q}_i$ (included the initial time $t^{q}_1$) need not
to follow any particular prescription, and are simply related to
Alice's source's repetition rate.

\noindent (5)~\textit{Transmission of classical information}. At
time $t^{b}_1 = (t^{q}_1+\tau+\Delta)$, \underline{without waiting
for Bob's receipt}, Alice starts the transmission of the basis
bits $b_{i}$ using the classical channel. At the generic time
$t^{b}_{i}=t^{q}_i+\tau+\Delta$ she will send out the bit $b_{i}$.
We note that $\tau$ and $\Delta$ have been declared on the
authenticated channel during step \underline{1}. We also note that
the classical channel can be thought for simplicity and without
loss of generality as the same channel used for the qubits, for
example an optical fiber, but with an intense signal traveling in
it. This entails that the bit $b_i$ employs an additional time
$\tau$ to reach Bob. In any case the traveling time on the
classical channel is measured and declared during
step~\underline{1}.

\noindent (6)~\textit{Acquisition of classical information}. At
certain times $T_{i}$ Bob acquires the $W$ basis bits $b_{i}$, and
labels them as $B_{i}$. This step is very similar to Bob receiving
a normal telephone call: he records both the values of the
$B_{i}$'s and their times of arrival $T_{i}$. For what said at
point \underline{5}, the expected times of arrival are $T_{i} =
(t^{q}_{i}+2\tau+\Delta+\delta)$, where $\delta>0$ is a certain
unavoidable temporal delay due to the electronics of Bob's
apparatus.

\noindent (7)~\textit{Deterministic measurement}. As soon as the
values $B_{i}$ are available to Bob, he uses them to perform a
deterministic measure of the qubits. The timing of this new
measure is given by $(T_{i}+\delta^{\prime})\pm\varepsilon$, where
$\delta^{\prime}\geq\delta$ is another temporal delay, known to
Bob, due to the imperfectness of his apparatus and
$\varepsilon\ll\Delta$ is the temporal acquisition window of his
detectors. Bob labels the outcomes of this measure as $D_{i}$ and
builds up the data string $D$.

\noindent (8)~After Bob's public announcement of the losses the
users should share on average $4N$ pairs of correlated bits (if it
is not so they abort the transmission). Alice selects a subset of
$2N$ bits of $d$ and $2N$ bits of $b$ that will serve as a check
of Eve's interference, and tells Bob the addresses of the selected
bits. Bob selects the same addresses from the strings $B$ and $D$.

\noindent (9)~Alice and Bob announce on the classical channel (i)
the values of the selected $2N$ basis bits $b_i$ and $B_i$. If any
of them does not coincide they abort the whole transmission. (ii)
The times of arrival $t^{b}_{i}$ and $T_{i}$ corresponding to the
selected $2N$ pairs of bits from $b$ and $B$. If any of them does
not fulfill the relation $T_{i} = t^{q}_{i}+2\tau+\Delta+\delta$
within the experimental error they abort the transmission (note
that for this step the clocks of the users are assumed to be
synchronized). (iii) The values of the selected $2N$ pairs of
check bits from $d$ and $D$. If more than an acceptable number of
these values disagree, they abort the transmission.

\noindent (10)~Alice and Bob perform error correction and privacy
amplification on the remaining $2N$ bits to obtain $2M$ private
key bits ($M \leq N$).

\medskip

This version of Det-BB84, with the crucial point \underline{5},
removes the problem of Bob's receipt, relying more on the
classical communication. The main ingredient is a kind of
``postselected'' receipt by Bob: Alice transmits the information
about the basis without waiting for Bob's receipt, and Bob does
not send the receipt in the very moment he receives the photon.
Yet his final measurement will reveal whether the photon was there
at the expected time or not. Thus the main problem of a QND
measurement is removed at the roots. Another advantage of the
above protocol is that the storage time at Bob's site is reduced
from $2\tau$ (as discussed in Section~\ref{sec:BB84}) to $\tau +
\Delta$. This reduces considerably the losses due to the storage
in Bob's quantum memory.

\subsection*{Security of the scheme} \label{sec:security}

Apart from the removal of Bob's receipt of the qubits, Det-BB84 is
entirely equivalent to the protocol we described in
Section~\ref{sec:BB84}, which, in turn, has been shown to be
secure and equivalent to the original BB84 in~\cite{Bih05}. Hence,
our security analysis aims at showing the security of Det-BB84
against attacks based on the potential weakness created by the
Bob's receipt removal. It can also be seen as a new security
argument in the frame of ``sequential'' protocols for
QKD~\cite{Mor96},~\cite{Cabello00}.

For the moment we consider Bob's measuring apparatus as ideal, and
we do not include in the proof the experimental parameters
$\delta$, $\delta^{\prime}$ and $\varepsilon$ introduced above.
The attackable point of our protocol is the lack of a qubit
receipt from Bob to Alice. The risk is that Eve uses the disclosed
bases to measure the qubits without perturbing them. Any other
kind of eavesdropping is tantamount to Eve attacking a qubit just
as she would do against a normal BB84 system. In order to exploit
the basis information Eve must delay the qubit until the basis is
disclosed. But any variation of the basis time of arrival respect
to what declared in point \underline{1} is detected during the
check performed at point (ii) of \underline{9}; and any variation
of the values of the bases decided by Alice, e.g. through an
impersonation attack~\cite{Dusek99}, is detected during the check
of point (i) of \underline{9}.

Then assume that Eve controls opportunely the length of the
channel between Alice and Bob in order to intercept the qubit
$q_i$, wait for the basis information $b_i$, measure the qubit
without perturbing it, and forward it to Bob without being
detected. Let us examine the timing of the protocol: Alice
transmits the qubit $q_i$ at time $t^{q}_i$, and the basis
information at time
\begin{equation}\label{basistime}
    t^{b}_{i}=t^{q}_i+\tau+\Delta.
\end{equation}
Bob waits for the basis and deterministically measures the qubit
at $T_i = t^{q}_{i}+2\tau+\Delta$ (if this last relation is not
\textit{a posteriori} satisfied the protocol is aborted, according
to point (ii) of \underline{9}). It is easy to see that Eve would
go undetected only if she is able to do her attack without
changing the time signature represented by $T_i$. Therefore, since
the storage time of Bob's quantum memory is $\tau + \Delta$, Eve
must let the qubit enter Bob's station at time
\begin{equation}  \label{t'}
T_i - (\tau + \Delta) =t^{q}_{i}+2\tau+\Delta - (\tau + \Delta) =
t^{q}_{i}+ \tau
\end{equation}
to go undetected. But this time is always less than that at which
the basis is revealed (Eq.\ref{basistime}), as long as $\Delta
> 0$. In other words when Eve knows the basis from Alice she does not
get the qubit anymore.

Now let us discuss the experimental delays $\delta$,
$\delta^{\prime}$ and $\varepsilon$ of Bob's apparatus in relation
to the security issue. The crucial quantity is the parameter
$\Delta$: how big should it be to maintain the security of the
protocol? The quantity $\varepsilon$ represents a kind of
experimental error in determining the exact time of arrival of the
photons at Bob's site. For example when the BB84 is implemented
using weak pulses as a photon source $\varepsilon$ is the time
window of Bob's ``gated mode'' detectors (i.e. detectors which are
open only when a photon is expected to be there); otherwise, when
the photons are generated through the spontaneous parametric down
conversion, $\varepsilon$ is the time window of the coincidence
counts. In both cases typical values of $\varepsilon$ are less
than 10 ns. In order to maintain the security of our protocol, at
point \underline{7} of Det-BB84 we required that
$\varepsilon\ll\Delta$. But it descends from our security argument
that actually the condition $\varepsilon<\Delta$ is sufficient to
guarantee the security of the protocol. Hence, for all practical
purposes, we can set $\Delta = 10 \varepsilon \approx 100$~ns.

\section{Experimental issues} \label{sec:practice}

In this section we discuss the feasibility of Det-BB84. We
consider a fiber-based configuration with weak pulses as a photon
source. For the only purpose of comparison we make our proposal
very similar to the one-way scheme recently reported on
in~\cite{Gob04}. However it is straightforward to work out a
different setup, for example using the two-way plug-and-play
configuration~\cite{Muller1996,Muller1997}.

The experimental Det-BB84 requires a fast and precise
synchronization: fast enough to reduce Bob's storage time, and
precise enough to fulfill the security criteria. Specifically
points \underline{4}-\underline{7} require a precise
synchronization between the line carrying the quantum information
(the qubit) and the line carrying the classical information (the
basis).

All the QKD realizations known so far use three lines for
communication: the quantum channel, the timing channel, or
trigger, and the classical channel. In~\cite{Gob04} the quantum
channel is a pulsed attenuated laser at the wavelength of
$1550$~nm, the trigger is a pulsed bright laser at the wavelength
of $1300$~nm, which is used to synchronize the whole apparatus,
and the classical channel is the Internet, which is employed to
transfer the information about the bases and about error
correction and privacy amplification. Now it is plain that to
obtain the synchronization between the qubit and the basis
mentioned above one can not rely on the Internet, because it can
be unpredictably slow and random in the delivery of the TCP/IP
packets. The solution is to use the trigger pulse, already
synchronized with the quantum line, to convey also the basis
information. To do that one can for example modulate the intensity
of the trigger pulse: `low-pulse' can represent a `0', while
`high-pulse' can represent a `1'. Or it is possible to adequately
digitalize the signal.

\begin{figure}[tbp]
\includegraphics[width=14cm]{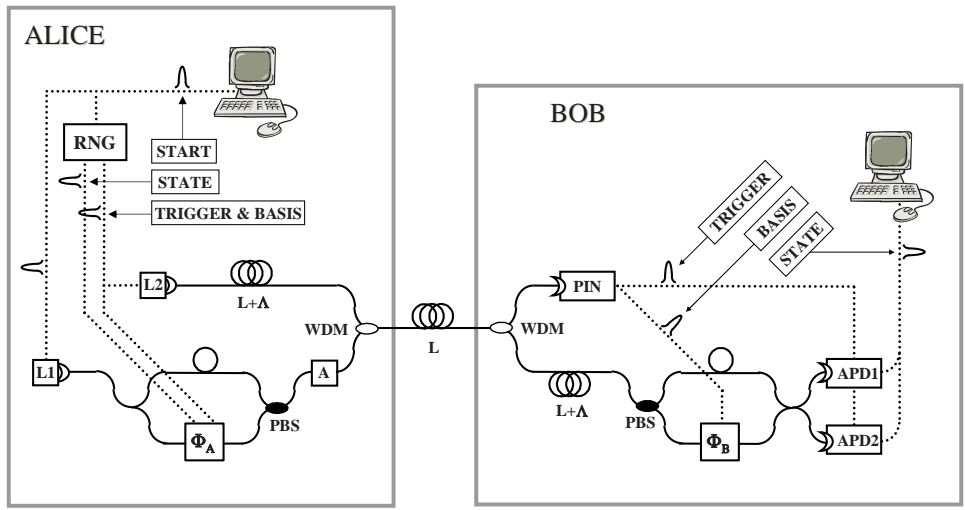}
\caption{Possible implementation of Det-BB84, obtained through
simple changes to the setup of Ref.\protect\cite{Gob04}. L1 and
L2: laser diodes; WDM: wavelength division multiplexer; PBS:
polarization beam combiner/splitter; APD: avalanche photo diode
detector. $L=\tau c/n$, $\Lambda=\Delta c/n$ (see text).}
\label{FIG:setup}
\end{figure}

Our scheme is sketched in Figure\ref{FIG:setup}.
The \textit{start} pulse from the computer drives the two laser
sources (L1 @1550 nm, the quantum signal, and L2 @1300 nm, the
trigger) and the phase modulator which encodes the information in
the relative phase of the two pulses generated by L1 and Alice's
interferometer. We drew the random number generator (RNG) as
detached from the computer for simplicity. The phase encoded on
the pulses is determined by the sum of the values of the basis
($0$ or $\pi/2$) and those of the state ($0$ or $\pi$). The
important feature is that the basis is also written on the bright
pulse @1300 nm, which now has a twofold role: time reference for
Bob and carrier of the basis information. Along the bright pulse
path there is a delay line, represented by a number of fiber
loops, of length $L+\Lambda$. To use the parameters given in
Det-BB84 we set $L=\tau c/n$, and $\Lambda=\Delta c/n$, with $n$
the refractive index of the fiber and $c$ the speed of light in
vacuum.

At Bob's site the WDM selects the bright pulse, which is directed
at a PIN photodiode detector. This acts as a trigger for the gate
of the avalanche photodiode detectors APD1 and APD2. Moreover the
value read by the detector (i.e. the basis used by Alice) acts as
an input to the phase modulator represented by $\varphi_B$ in the
figure, thus allowing the deterministic measurement by Bob. On the
other hand, the path followed by the quantum carrier (photon from
laser L1) is the same as in~\cite{Gob04}. The only difference is
the delay on Bob's site, which is equal to the one at Alice's.
This delay represents the simplest quantum memory and allows Bob
to wait for the information about the basis before his final,
deterministic measurement. So in the whole, with respect to the
usual BB84, no additional material other than some software and
electronics is required for the implementation of Det-BB84.

\subsection*{Comparison with the BB84} \label{sec:comparison}

In this section we compare our proposal for a practical Det-BB84
with the BB84 of~\cite{Gob04} in terms of the rate of secure bits,
$R_{sec}$, introduced by N. Lutkenhaus in~\cite{Lutkenhaus2000}.
$R_{sec}$ is a pure number and represents the fraction of
distilled secure bits after the procedures of error
correction~\cite{Bennett1992,Brassard1994,But03} and privacy
amplification~\cite{Bennett1995}. It must be multiplied by the
effective repetition rate in order to obtain the total secure rate
of the considered setup. The $R_{sec}$ for a BB84 implemented with
weak pulses is defined as~\cite{Lutkenhaus2000}:
\begin{equation}  \label{R_BB84}
R_{sec}^{BB84}=\frac{1}{2} p_{exp} \{
\beta\left[1-\tau\left(e/\beta\right)\right]-
f_{casc}h\left(e\right) \}.
\end{equation}
The coefficient $1/2$ comes from the basis reconciliation
procedure, in which the users' bases coincide with an average
probability of $1/2$~\cite{note3_DetBB84}. $p_{exp}$ is the signal
of the experiment, which is given by the formula:
\begin{equation}\label{pexp}
    p_{exp}=p_{exp}^{signal}+p_{exp}^{dark}-p_{exp}^{signal}p_{exp}^{dark};
\end{equation}
$p_{exp}^{dark}$ is the probability Bob gets a dark count in his
detectors, while $p_{exp}^{signal}$ is the probability that Bob's
detector fires because of a photon emitted by Alice's source. This
probability decreases with the distance between the users
according to the expression:
\begin{equation}\label{pexpsignal}
    p_{exp}^{signal}=1-exp\left( -\eta_{B}\eta_{T}\mu \right),
\end{equation}
where $\eta_{B}$ is the quantum efficiency of Bob's detectors,
$\mu$ is the average number of photons per pulse, and $\eta_{T}$
is the transmission probability of the channel, given by:
\begin{equation}\label{etaT}
   \eta_{T}=10^{-\left(\alpha L + L_{c}    \right)/10}.
\end{equation}
$\alpha$ is the absorption coefficient of the fiber, $L_{c}$ is
the loss rate at receiver's station and $L$ is the distance
between the users, as reported in Figure~\ref{FIG:setup}.
Furthermore in Eq.~(\ref{R_BB84}) $\beta$ is defined as:
\begin{equation}\label{beta}
    \frac{p_{exp}-S_{m}}{p_{exp}},
\end{equation}
with $S_{m}$ the probability that Alice photon source emits more
than a single photon per pulse. $\beta$ is a sort of security
parameter: until it is positive the protocol is secure against the
so-called PNS attacks~\cite{Hut95,Lutkenhaus2000,Bra00}.
$f_{casc}$ is a function defined in~\cite{Brassard1994} that takes
into account the imperfect (although efficient) error correction
procedure performed with the Cascade protocol. For simplicity we
set it equal to $1$ in our simulations. $h\left( e\right) $ is the
Shannon entropy pertaining to a given QBER $e$. Finally $\tau$ is
the fraction of the error-corrected key which has to be discarded
during privacy amplification when only single-photon pulses are
taken into account~\cite{Lutkenhaus1999}; it is a function of the
QBER and amounts to: $\tau(e) =\log_{2}(1+4e-4e^{2})$ for $0\leq
e\leq1/2$ and $\tau(e) =1$ for $1/2<e\leq1$.

\smallskip

Analogously we define the secure rate for Det-BB84 as:
\begin{equation}  \label{R_DetBB84}
R_{sec}^{Det-BB84}=p_{exp} \{
\beta\left[1-\tau\left(e/\beta\right)\right]-
f_{casc}h\left(e\right) \}.
\end{equation}
Notice that the coefficient $1/2$ is replaced by $1$ in the above
equation, due to Bob's deterministic measurement. Furthermore the
transmission probability $\eta_{T}$ is different from the one in
Eq.~(\ref{etaT}). In fact in our scheme the photon is stored in
the fiber loops at Bob's site, whose length is $L + \Lambda$. Then
the transmission probability becomes:
\begin{equation}\label{etaT_DetBB84}
   \eta_{T}^{\prime}=10^{-\left[\alpha \left(2L + \Lambda \right) + L_{c}    \right]/10}.
\end{equation}
This entails that Det-BB84 is more affected by losses than BB84.
However when the distance between Alice and Bob is small enough,
the loss-rate is low, and the determinism still provides a
nontrivial increase of the secure bit-rate.

For every fixed distance $L$ between the users the secure rate has
a different maximum in the average photon number
$\mu$~\cite{Lutkenhaus2000}. In our numerical simulation we chose
the value of $\mu$ as such as to independently maximize the secure
rate of BB84 and Det-BB84 at given lengths $L$. These values are
reported in Table~\ref{tableI}.
\begin{table}[h!]
  \centering
\begin{tabular}{|c|c|c|c|c|c|}
  \hline
  distance (km) & $\mu_{optimal}^{Det-BB84}$ & $\mu_{optimal}^{BB84}$ & $R_{\text{sec}}^{Det-BB84}$ & $R_{\text{sec}}^{BB84}$ & $R_{\text{sec}}^{Det-BB84}/R_{\text{sec}}^{BB84}$ \\
  \hline
  2 & 0.03820 & 0.04200 & 6.9145$\times10^{-4}$ & 4.1686$\times10^{-4}$ & 1.6587 \\
  4 & 0.03155 & 0.03818 & 4.7567$\times10^{-4}$ & 3.4572$\times10^{-4}$ & 1.3759 \\
  8 & 0.02162 & 0.03156 & 2.2498$\times10^{-4}$ & 2.3783$\times10^{-4}$ & 0.9460 \\
  16 & 0.01025 & 0.02165 & 4.9456$\times10^{-5}$ & 1.1249$\times10^{-4}$ & 0.4396 \\
  \hline
\end{tabular}
  \caption{Values used for the numerical calculation of the secure key rates of
  BB84 and Det-BB84 protocols. For each distance and for each protocol the average
  photon number $\mu$ has been optimized to maximize the rate. Experimental
parameters taken from~\cite{Gob04}.}\label{tableI}
\end{table}

\begin{center}
\begin{figure}[h!]
\includegraphics[width=14cm,height=10cm]{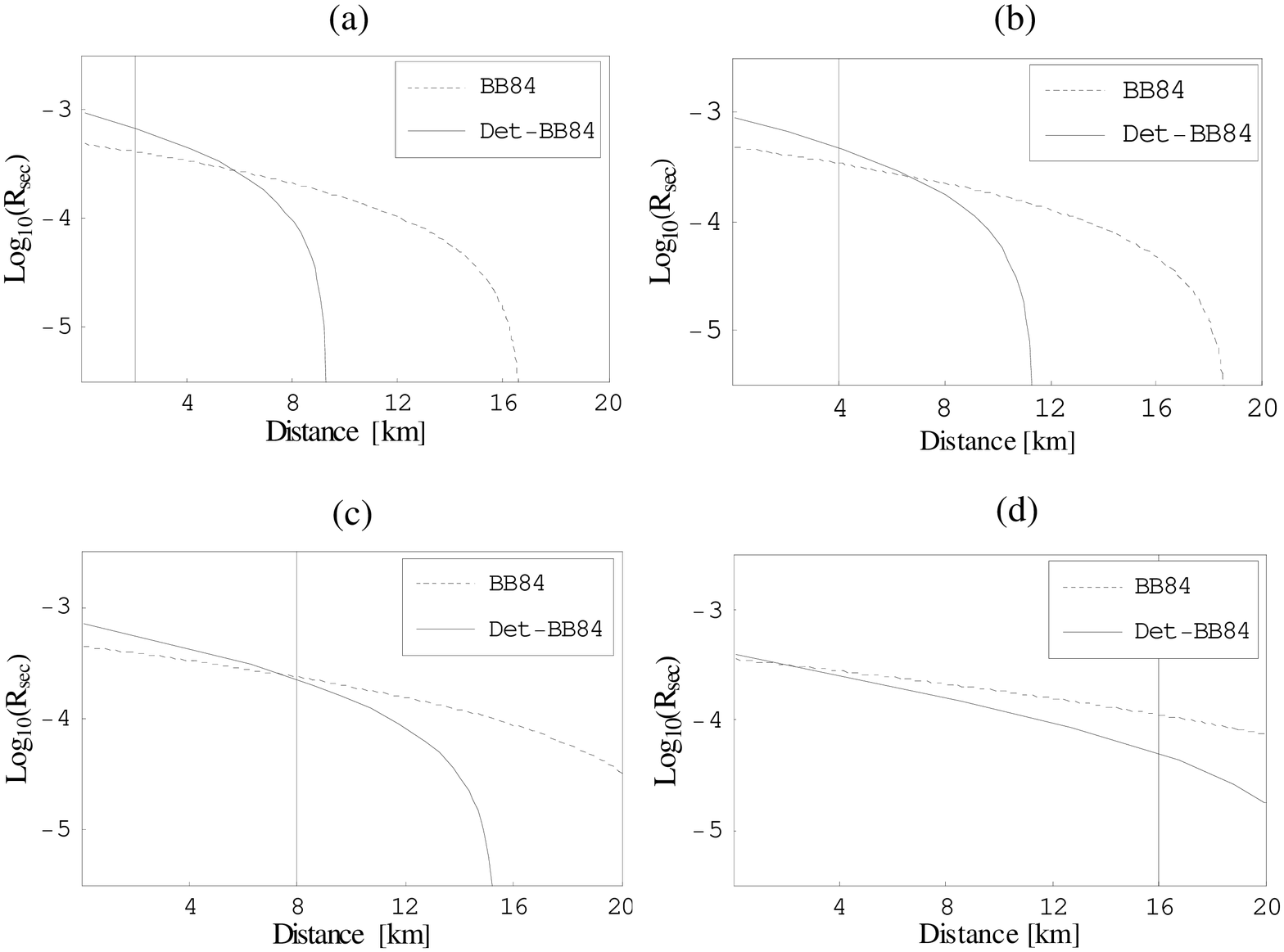}\newline
\caption{Secure rate of Det-BB84 and BB84 optimized for distances
between Alice and Bob of 2, 4, 8 and 16 km. Experimental
parameters taken from~\cite{Gob04}.} \label{FIG1}
\end{figure}
\end{center}

In Figure~\ref{FIG1} the secure rate is plotted for BB84
(Eq.~\ref{R_BB84}) and Det-BB84 (Eq.~\ref{R_DetBB84}) as a
function of $L$. Only the average photon number $\mu$ is
different, according to what just explained. The diagrams
(\textit{a}), (\textit{b}), (\textit{c}) and (\textit{d}) have
been obtained by fixing four values of $L$ and finding the values
$\mu_{i}$ that maximize $R_{sec}\left(\mu_{i}|L_{j}\right)$
separately for BB84 and Det-BB84. Vertical lines have been drawn
at the crucial distances $L_{j}$.

It can be seen that in the plots (\textit{a}) and (\textit{b}) the
secure rate provided by the Det-BB84 is higher than that
pertaining to BB84. After that, in plot (\textit{c}), the rates
provided by the two protocols are almost the same. Finally in plot
(\textit{d}), the standard BB84 provides a higher rate. In other
words, for distances up to about 8 km the Det-BB84 provides a
better rate than the non-deterministic BB84. For distances of less
than 2 km the improvement factor is more than 1.65, nearing the
final value of 2 for very short distances and for a lossless
setup. We remember that the maximum secure distance achievable
with the BB84 setup described in~\cite{Gob04} is about 60 km.

\medskip

It should be noted that the secure rate is a figure of merit of a
QKD setup, and is not a trivial task to increase it. The rate of
transmission in any fiber-based setup is currently limited by
detectors' minimum dead times, which are of the order of
microseconds for a standard InGaAs Avalanche Photodiode Detector
(APD). This is a technological limitation that can be surpassed
only by improving the detection mechanism. All the same, in the
setups exploiting the spontaneous parametric down-conversion as a
single-photon source is not possible to increase the signal on
demand. The improvement brought about by Det-BB84 works in both
the situations as it concerns the protocol itself, not the way it
is implemented. In this respect the plots in Figure~\ref{FIG1} are
``universal'', i.e. independent of the particular technology
employed in the experiments. For example it is possible to simply
change the scaling factor of the plots drawn above to know the
performances of a Det-BB84 realized in free space at the
wavelength of 800 nm. It is worthwhile to mention that a low-loss
high-rate QKD on very short distances has attracted recently
renewed interest because of its closeness to the credit-card
security issue~\cite{Dul06}. The rate of this kind of
transmissions can be almost doubled using Det-BB84.

We also remark that the performances of the practical Det-BB84
studied here are not the best possible. For instance we assumed a
poor quantum memory for the storage of the photons at Bob's site,
i.e. an optical-fiber loop with the same transmission as the one
used to connect the users. The maximum distance over which the
Det-BB84 outperform the BB84 directly depends on this storage
mechanism: the better it is the longer the distance. Technological
progresses concerning quantum memories are continuously reported,
and values of storage time up to microseconds have been recently
achieved~\cite{storage}.

\section{Conclusion}

In this paper we have provided a first evidence that a
deterministic rendering of the BB84 protocol is not only
conceivable in theory, but also realizable in practice with
current technology. We have introduced the new protocol Det-BB84
which is as secure as the standard BB84 and does not need a Bob's
receipt for the qubits, thus paving the way to a feasible setup.
In fact we have proposed an implementation of Det-BB84 and
compared it with the BB84 reported in~\cite{Gob04} in terms of the
secure rate of distilled bits, finding a nontrivial increase of
the bit-rate at small distances between Alice and Bob. This
increase can become even more relevant as long as technology
provides better memories for the storage of the qubits.

\section{Acknowledgements}

The core of this work has been prepared in the labs of MIMOS
Berhad, Kuala Lumpur. For its completion one of the authors (M.
L.) acknowledges the European Commission through the Integrated
Project ''Qubit APplications'' (QAP), Contract No. 015848, funded
by the IST directorate.

\newpage

\bibliography{total}

\begin{thebibliography}{48}
\bibitem[{not({\natexlab{b}})}]{note3_DetBB84}
\bibinfo{note}{It should be noted that an efficient BB84, featuring 1 as a
  coefficient, has been proposed in~\cite{Lo05}. However that scheme is only
  asymptotically efficient, and, in order to mantain the unconditional
  security, it foresees a number of check bits increasing as $W/p^{2}$, with
  $p$ the bias of the bases. Our scheme is conceptually and practically
  different.}
\end{thebibliography}

\end{document}